\documentclass[aps,pre,twocolumn,showpacs,amsmath,amssymb]{revtex4}
\usepackage{graphicx}
\usepackage{epsf}
\usepackage{bm}

\def\be{\begin{equation}}
\def\en{\end{equation}}
\newcommand{\bi}[1]{\mbox{\boldmath$#1$}}

\begin{document}
\title{
Efficient simulations of charged colloidal dispersions: A density
functional approach
}
\author{Kang Kim$^1$}

\author{Ryoichi Yamamoto$^{1,2}$}

\affiliation{
$^{1}$PRESTO, Japan Science and Technology Agency, 4-1-8 Honcho
Kawaguchi, Saitama, Japan
}
\affiliation{
$^2$Department of Physics, Kyoto University, Kyoto 606-8502, Japan
}
\date{\today}

\begin{abstract}
A numerical method is presented for first-principle simulations of charged 
colloidal dispersions in electrolyte solutions.
Utilizing a smoothed profile for colloid-solvent boundaries, 
efficient mesoscopic simulations are enabled for modeling 
dispersions of many colloidal particles exhibiting many-body
electrostatic interactions.
The validity of the method was examined for simple colloid geometries, 
and the efficiency was demonstrated by calculating stable structures of 
two-dimensional dispersions, which resulted in the formation of
colloidal crystals.
\end{abstract}

\maketitle

\section{introduction}

Electrostatic interactions play a crucial role in 
colloidal dispersions~\cite{Israelachvili,Russel,Safran}.
When colloidal particles are immersed in electrolyte solutions,
the so-called electric double layer is formed.
The electric double layer is a cloud of 
counterions dissociated from the surfaces of the colloids into the
solvent surrounding the colloidal particles.
Most counterions are localized within the double layer 
due to the electrostatic attractive interactions between counterions
and the inversely charged colloidal surfaces,
while the entropy of the counterions tends to delocalize them.
The thickness of the double layer is then determined by the 
competition between these conflicting effects.
The static density profiles of the counterions can be calculated 
properly using the Poisson--Boltzmann theory, and its linearized version 
leads to the well-known screened Coulomb interaction for a pair of 
likely charged colloids.
Although the screened Coulomb potential is widely used to simulate 
charged colloidal dispersions, the linearization is justified only 
for large interparticle separations.
Deviations from screened Coulomb interaction become
notable for the interparticle separation smaller than the Debye
screening length.
Furthermore, many-body interactions become significant for dense 
colloidal dispersions.
We thus need an alternative framework which is applicable for 
simulating dense dispersions composed of many charged colloids.

In principle, the above problems can be resolved properly if 
molecular dynamics (MD) or Monte Carlo (MC)
simulations are used with treating counterions 
explicitly as millions of charged particles.
From a computational point of view, however, such fully microscopic 
simulations are prohibitively inefficient because of the huge 
asymmetries both in size and time 
scales between colloidal particles (large and slow) and counterions 
(small and fast).
An enormous number of counterions and simulation steps are 
required even for a system composed of only a few colloidal particles.
Alternatively, counterions can be modeled as a coarse-grained continuum
object which is governed by a set of partial differential equations (PDEs)
with appropriate boundary conditions defined at the fluid-colloid
interface as demonstrated in some previous studies 
\cite{M_Fushiki1992,J_Dobnikar2003EPL,J_Dobnikar2003JCP}
rather than microscopic particles governed by 
Newton's equations of motion.
This idea can be most simply implemented by utilizing the 
finite element method (FEM), which is a very natural and sensible 
method to simulate solid particles with arbitrary shapes in a
discrete computational space.
Several boundary-fitted unstructured mesh have been applied to specific
problems, so that the shapes of the particles are properly
expressed~\cite{W_R_Bowen1998,P_Dyshlovenko2000,C_Russ2002}.
Thus, in principle it is possible to apply this method to
dispersions consisting of many particles with any shape.
However, a numerical inefficiency arises from the following:
i) re-constructions of the irregular mesh are necessary at every
simulation step according to the temporal particle positions, and ii)
PDEs must be solved under boundary conditions imposed on the surfaces 
of all colloidal particles.  
The computational demands thus are enormous for systems involving many
particles, even if the shapes are all spherical.

In order to overcome the problems mentioned above, 
a new idea was put forwarded by L{\"o}wen {\it et al.} 
for their first principle MD simulations of charged colloidal dispersions
\cite{H_Lowen1992,H_Lowen1993JCP,H_Lowen1993EPL,H_Lowen1994,R_Tehver1999}.
Utilizing a pseudopotential, this method enabled us to use 
the conventional Cartesian coordinate 
and to calculate the force acting on each particle
due to the solvents efficiently.
In the present paper, we extend this idea by introducing 
a smoothed profile function $\phi({\bf r})$ to represent the 
colloid-solvent interface with a finite thickness $\xi$.
The validity of the method is examined in some simple situations 
with changing the interface thickness $\xi$.
Then the efficiency of the method is demonstrated by simulating 
two-dimensional dispersions, which resulted in the formation of
colloidal crystals.
The same type of method has already been applied successfully 
to liquid-crystal colloid dispersions \cite{R_Yamamoto2001,R_Yamamoto2004}.

\section{density functional approach}

\subsection{original governing equations}

According to the density functional theory established for charged 
colloidal dispersions~\cite{Safran,Barrat,JP_Hansen2000},
colloids are treated explicitly as particles, while counterions 
are treated as a continuum object.
Let us consider a system consisting $N$ negatively charged colloidal 
particles with radius $a$ and a solution of monovalent counterions and
coions whose bulk concentrations are set to $\rho_0$.
Each colloidal particles is carrying a negative charge $-Ze$,
which is distributed uniformly on its surface of area $A=4\pi a^2$
(three-dimension) or $=2\pi a$ (two-dimension).
Here $e$ represents the unit charge.
The solvent is assumed to have an uniform dielectric constant
$\epsilon$, and spatial distributions of counterions($+$) and
coions($-$) are characterized by the local number density 
$\rho_{+}({\bi r})$ and $\rho_{-}({\bi r})$. 
The overall charge neutrality of the system is guaranteed by the
constraint
\begin{equation}
\int d{\bi r} e\rho_{c}({\bi r})= NZe
\label{neutral}
\end{equation}
with $e\rho_c({\bi r})\equiv e(\rho_{+}({\bi r})-\rho_{-}({\bi r}))$.
The integral runs over the total volume and $\rho_{\pm}({\bi r})=0$ is
to be strictly satisfied if ${\bi r}$ is inside of the particles.
For a given colloidal configuration $\{{\bi R}_1,\cdots,{\bi R}_N\}$,
the free energy ${\mathcal F}$ of the system is given
by the functional of $\rho_{\pm}({\bi r})$ as~\cite{Safran,Barrat,JP_Hansen2000}
\begin{equation}
{\mathcal F}[\rho_{\pm}({\bi r}); {\bi R}_1,\cdots,{\bi R}_N]
\equiv
{\mathcal F}_{id}+{\mathcal F}_{ele},
\label{free}
\end{equation}
\begin{eqnarray}
{\mathcal F}_{id}&=& k_BT \sum_{\alpha=+,-}\int d{\bi r}
\rho_{\alpha}({\bi r})\left\{\ln\left[\frac{\rho_{\alpha}({\bi
r})}{\rho_0}\right]-1\right\},
\label{freeid}\\
{\mathcal F}_{ele}&=&\frac{1}{2} \int d{\bi r}
e[\rho_{c}({\bi r})+q({\bi r})]\psi({\bi r}),
\label{freeele}
\end{eqnarray}
where ${\mathcal F}_{id}$ and ${\mathcal F}_{ele}$
represent the ideal gas contribution of the ions and
the electrostatic contribution resulting from Coulomb
interaction, respectively.
The distribution of the surface charge of colloidal particles is given by
\begin{equation}
eq({\bi r})=-\frac{Ze}{A}\sum_{i=1}^{N}\delta(a-|{\bi r}-{\bi R}_i|),
\label{charge}
\end{equation}
and $\psi({\bi r})$ is the electrostatic potential
defined by the solution of the Poisson equation
\begin{equation}
\nabla^2 \psi({\bi r}) = -\frac{4\pi e}{\epsilon}[\rho_{c}({\bi r})+q({\bi r})].
\label{poisson}
\end{equation}

The equilibrium density of counterions and coions 
$\rho_{\pm}^{eq}({\bi r})$ are given by the solution of the variational
equation
\begin{equation}
\left.\frac{\delta \Omega}{\delta \rho_{\pm}({\bi r})}\right|_{\rho_{\pm}({\bi r})
=\rho_{\pm}^{eq}({\bi r})}=0,
\label{variational}
\end{equation}
where $\Omega\equiv {\mathcal F} - \mu \int d{\bi r} \rho_c({\bi r})$
is the grand potential functional and $\mu$ is the chemical potential.
Equation (\ref{variational}) leads to 
\begin{equation}
\rho_{\pm}^{eq}({\bi r})=\rho_0 \exp[(\mp e \psi({\bi r})\pm \mu)/k_BT],
\label{boltzmann}
\end{equation}
where the constant $\mu$ must be determined by the charge neutrality
condition Eq.~(\ref{neutral}).
Substituting $\rho_{\pm}^{eq}({\bi r})$ for $\rho_{\pm}({\bi r})$ in
Eq.~(\ref{poisson})
yields the Poisson--Boltzmann (PB) equation.
From a computational point of view, solving the PB 
equation for dispersions of many colloidal particles is 
quite demanding because the equation must be solved 
iteratively to impose correct 
boundary conditions defined on surfaces of all the colloidal particles. 
Usually, this can be done by employing non-Cartesian coordinate
systems as in FEM~\cite{W_R_Bowen1998,P_Dyshlovenko2000,C_Russ2002},
which however makes numerical calculations very complicated and
inefficient for dispersions involving many moving particles.
Another serious problem arises if one calculates
the force acting on colloids induced by the counterions 
${\bi f}_i^{PS}=-\partial\Omega/\partial {\bi R}_i$
due to singularities in ${\mathcal F}$ similar to the case of
liquid-crystal colloid dispersions
\cite{R_Yamamoto2001,R_Yamamoto2004}.

\subsection{smoothed profile method}

In order to improve the numerical inefficiency due to the moving
boundary problem, a smoothed profile was 
introduced to the colloid-solvent 
interface with its thickness $\xi$ rather than the original
sharp profile corresponds to $\xi=0$.
This simple modification greatly benefits
the performance of numerical computations.
Similarly to earlier studies 
\cite{R_Yamamoto2001,R_Yamamoto2004,Y_Nakayama2004,H_Tanaka2000,H_Kodama2004},
the smoothed profile function
\begin{equation}
\phi_i({\bi r})=\frac{1}{2}
\left[ \tanh\left(\frac{a-|{\bi r}-{\bi R}_i|}{\xi}\right)+1 \right],
\label{interface}
\end{equation}
is used for the individual particle $i$, where $\xi$ represents the 
interface thickness.
Then, the governing equations (\ref{neutral}) and 
(\ref{freeid})-(\ref{poisson}) can be re-written using 
the overall profile function $\phi({\bi r})=\sum_{i=1}^N \phi_i({\bi r})$ as
\begin{equation}
\int d{\bi r} \left(1-\phi({\bi r}) \right)e\rho_c({\bi r}) = NZe,
\label{neutral2}
\end{equation}
\begin{eqnarray}
{\mathcal F}_{id}
&=&k_BT\sum_{\alpha=+,-}\int d{\bi r} \left(1-\phi({\bi r}) \right)
\rho_{\alpha}({\bi r})\left\{\ln\left[\frac{\rho_{\alpha}({\bi r})}{\rho_0}\right]-1\right\},
\label{free2id}
\\
{\mathcal F}_{ele}
&=&\frac{1}{2}\int d{\bi r} e\left[ \left(1-\phi({\bi r}) \right)
\rho_c({\bi r})+ q({\bi r})\right] \psi({\bi r}),
\label{free2ele}
\end{eqnarray}
\begin{equation}
eq({\bi r})=-\frac{Ze}{A} \sum_{i=1}^N |\nabla\phi_i({\bi r})|,
\label{charge2}
\end{equation}
\begin{equation}
\nabla^2 \psi({\bi r}) = -\frac{4\pi e}{\epsilon}
\left[\left(1-\phi({\bi r}) \right)
\rho_c({\bi r})+q({\bi r})\right].
\label{poisson2}
\end{equation}
The delta function in Eq.~(\ref{charge}) is replaced with a smooth
function $|\nabla\phi_i({\bi r})|$ in Eq.~(\ref{charge2}) to remove
the numerical singularity.
Since $\rho_c({\bi r})$ and $q({\bi r})$ are now continuous functions of
${\bi r}$ in the whole domain without any boundary conditions,
the electrostatic potential
$\psi({\bi r})$ can be obtained very efficiently by using the fast Fourier
transform to solve Eq.~(\ref{poisson2}). 
The equilibrium density profile $\rho_{\pm}^{eq}({\bi r})$ is calculated
iteratively until Eqs.~(\ref{boltzmann}), (\ref{neutral2}), and
(\ref{poisson2}) become self-consistent.
The grand potential is given by 
\begin{equation}
\Omega[\rho_{\pm}({\bi r}); {\bi R}_1,\cdots,{\bi R}_N]={\mathcal F}-\mu \int
d{\bi r} \left(1-\phi({\bi r}) \right)\rho_c({\bi r}).
\end{equation}
We note that the above equations Eqs.~(\ref{neutral2})-(\ref{poisson2})
reduce simply to the original Eqs.~(\ref{neutral}) and 
(\ref{freeid})-(\ref{poisson}), respectively, for $\xi \to 0$.

Once the equilibrium density $\rho_{\pm}^{eq}({\bi r})$ is determined, 
the force ${\bi f}_i^{PS}$ acting on the $i$th particle 
directly follows the Hellmann--Feynman theorem,
\begin{eqnarray}
&&{\bi f}_i^{PS}(\rho_{\pm}^{eq}({\bi r});{\bi R}_1,\cdots,{\bi R}_N)=
-\frac{\partial \Omega[\rho_{\pm}^{eq}({\bi
r}); {\bi R}_1,\cdots,{\bi R}_N]}{\partial {\bi R}_i}\nonumber\\
&&=k_BT\sum_{\alpha=+,-}\int d{\bi r} \frac{\partial \phi_{i}({\bi r})}{\partial {\bi R}_i}
\rho_{\alpha}^{eq}({\bi r})\left\{\ln \left[\frac{\rho_{\alpha}^{eq}({\bi r})}{\rho_o}\right]-1\right\}\nonumber\\
&&-\int d{\bi r} e\left[ -\frac{\partial \phi_{i}({\bi r})}{\partial {\bi
R}_i}\rho_c^{eq}({\bi r})+\frac{\partial q({\bi r})}{\partial {\bi
R}_i}\right]\psi({\bi r})\nonumber\\
&&-\mu\int d{\bi r} \frac{\partial \phi_{i}({\bi r})}{\partial {\bi
R}_i}\rho_c^{eq}({\bi r}).
\label{HF}
\end{eqnarray}
One can easily calculate the force since both 
$\partial \phi_{i}({\bi r})/\partial {\bi
R}_i$ and $\partial q({\bi r})/\partial {\bi R}_i$ are 
analytical functions of ${\bi R}_i$.
The particle positions should be updated using the appropriate
equations of motion, 
\begin{eqnarray}
m_i\frac{d^2 {\bi R}_i}{dt^2}={\bi f}_i^{PP}+{\bi f}_i^{PS}+{\bi f}_i^{H}+{\bi f}_i^{R},
\label{motion}
\end{eqnarray}
where ${\bi f}_i^{PP}$ represents the force due to direct
particle-particle interactions, and ${\bi f}_i^{H}$ and ${\bi
f}_i^{R}$ are the hydrodynamic and random forces.
Repeating this procedure enables us to perform
first-principle simulations for charged colloidal dispersions
without neglecting many-body interactions.

The present method can be regarded as an alternative representation 
of the idea proposed by 
L{\"o}wen {\it et al.} 
\cite{H_Lowen1992,H_Lowen1993JCP,H_Lowen1993EPL,H_Lowen1994,R_Tehver1999}, 
where
the total charge of a single particle is assumed to be located at the
center of the particle and the penetration of counterions into the
colloids are avoided effectively by using a pseudopotential between 
colloids and counterions.
Kodama {\it et al.} extended this idea in their recent paper
\cite{H_Kodama2004}, where the total charge is distributed on 
the colloidal surface and the pseudopotential is used to avoid 
the penetration. 
In the present method, the densities of counterions and coions are 
defined as $(1-\phi({\bi r}))\rho_{\pm}({\bi r})$ and 
the total charge of a colloid is distributed on its surface by using 
also the profile function $\phi({\bf r})$
so that the ions cannot penetrate into colloids explicitly 
and the conservation of the ions are satisfied automatically.

\section{numerical calculations}

We have performed simulations for two-dimensional dispersions 
composed of colloidal disks, 
which are the two-dimensional representation of infinitely 
long cylindrical rods.
The Debye screening length is given by
$\lambda_D=1/\sqrt{8\pi\lambda_B \rho_0}$, where
$\lambda_B=e^2/\epsilon k_BT$ is the Bjerrum length and used as 
the unit of length in this paper.
The radius and line charge density at the surface of the 
disks are chosen as
$a=5$ and $\lambda e\equiv Ze/\lambda_B=e$, respectively.

\begin{figure}
\includegraphics[scale=0.3]{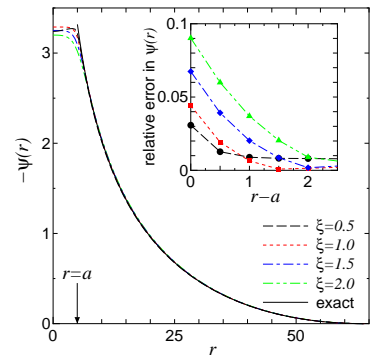}
\caption{
\label{psi}
Electrostatic potential $\psi(r)$ around an infinitely long cylindrical
rod of radius $a=5$ as a function of the distance from the center.
The solid line indicates the analytic solution.
The electrostatic potential and the length are measured in units of
$k_BT/e$ and $\lambda_B$, respectively.
Inset: relative error in the electrostatic potential for
$\xi=0.5$ (circles), $1.0$ (squares), $1.5$ (diamonds), and $2.0$
(triangles) as a function of $r-a$.
}
\end{figure}

To test the accuracy of the present method,
we first consider a rod located at a center of the cylindrical 
(Wigner--Seitz) cell whose inner radius is $R=64$.
Here we assume the system contains no additional salt.
The analytical solution of the PB equation is known for this simple
geometry, and the electrostatic potential is 
given by~\cite{RM_Fuoss1951,T_Alfrey1951}
\begin{equation}
\psi(r)=\ln\left[\frac{(\kappa_D
r)^2}{R_m}\cos^2\left(\gamma\ln\left[\frac{r}{R_m}\right]\right)\right],
\label{PB_exact}
\end{equation}
for $a<r<R$.
Here, the constant $\gamma$ is determined as the solution of the
algebraic equation
\begin{equation}
\tan^{-1}\left(\frac{\lambda-1}{\gamma}\right)+\tan^{-1}\left(\frac{1}{\gamma}\right)-\gamma\ln\left(\frac{R}{a}\right)=0,
\label{gamma}
\end{equation}
where the constant $R_m$ is given by
\begin{equation}
R_m=R\exp\left[\frac{1}{\gamma}\tan^{-1}\left(\frac{\lambda-1}{\gamma}\right)\right].
\label{R_m}
\end{equation}
Furthermore, the inverse screening length 
$\kappa_D$
is defined by
$\kappa_D^2=4(1+\gamma^2)/R$.
The electrostatic potential is calculated using our method 
with $\xi=0.5$, $1.0$, $1.5$, and $2.0$ and plotted in Fig.~\ref{psi}.
We see that the numerical results are in good
agreements with the analytic solution for $r-a>\xi$, though
some deviations are found for $r-a \leq \xi$ as an
artifact of the smoothed profile.
We emphasize that the inset of Fig.~\ref{psi} shows 
the relative error $(\psi(r,\xi)-\psi_{\rm exact}(r))/\psi_{\rm exact}(r)$ 
is only within $1\%$ for $r-a> \xi$. 

\begin{figure}
\includegraphics[scale=0.3]{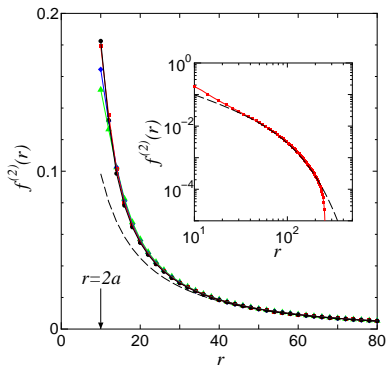}
\caption{
\label{2force}
The pair force $f^{(2)}(r)$ between two colloidal disks as a function of 
the inter-disk separation $r$ with $\xi=0.5$ (circles), $1.0$
(squares), $1.5$ (diamonds), and $2.0$ (triangles).
The dashed line indicates the LPB solution.
The units of force and length are $k_BT/\lambda_B^{2}$ and $\lambda_B$, 
respectively.
Inset: log-log plot of $f^{(2)}(r)$.
}
\end{figure}

We next consider the pair interaction between charged 
disks immersed in a solution of counterions and coions. 
The system consists of $1024\times 1024$ grid points with the periodic
boundary condition.
The linear length of the system is $L=512$ and 
the Debye screening length is chosen as $\lambda_D=50$
in unit of $\lambda_B$.
The equilibrium density profile is calculated for given
inter-disk separations $r$, then, the force acting on the pair 
$f^{(2)}(r)$ is calculated using Eq.~(\ref{HF}) for
different values of the interface thickness $\xi=0.5$, $1.0$, $1.5$, and $2.0$.
In Fig.~\ref{2force}, the force obtained by the present method is
plotted and compared with the analytical solution 
of the linearized PB (LPB) equation~\cite{JP_Hansen2000}, 
$f_{LPB}^{(2)}(r)=-dv/dr$ with
\begin{equation}
v(r)=\frac{(\lambda e)^2}{(\lambda_D^{-1}a K_1(a/\lambda_D))^2\epsilon}
K_0(r/\lambda_D),
\label{LPB}
\end{equation}
where $K_0(x)$ and $K_1(x)$ are Bessel functions of imaginary argument.
Since the thickness of the electric double layer is roughly given by
the Debye screening length $\lambda_D$,
the force obtained by our numerical method agrees well with the
linearized solution $f_{LPB}^{(2)}(r)$ for $r-2a > \lambda_D$.
For short distances $r-2a < \lambda_D$, on the other hand, deviations
from the LPB solution 
become notable.
It is seen in the inset of Fig.~\ref{2force} that the
deviations between numerical results and the LPB solution become 
notable also for large $r$.
This is nothing more than the artifact of the periodic boundary 
condition used in our numerical 
calculations where $f^{(2)}(r)$ must be zero at $r=L/2=256$.

The dependence of the interface thickness $\xi$ on
$f^{(2)}(r)$ is similar to the case of he electrostatic potential
$\psi(r)$ shown in Fig.~\ref{psi}.
The all numerical curves in Fig.~\ref{2force} are almost identical 
(within $1\%$ error) for $r-2a<2\xi$, 
while some deviations are observed for very short distances $r-2a\leq 2\xi$.
In this case, the overlapping of the interface functions $\phi_i({\bi r})$
occurs between two disks.
If we use a very small $\xi$ and an infinite number of grid points,
we may reproduce the force curve obtained by more accurate, 
but numerically more expensive, methods like FEM quantitatively 
for small inter-disk separations.
However, the numerical cost would also be very expensive in such a case.
In other words, the trade-off for the increase in numerical efficiency 
using the non-zero interface thickness is some loss of the numerical 
accuracy.
This may give rise to some quantitative errors when the separation 
between colloids are very small $r-2a\leq 2\xi$.

\begin{figure}
\includegraphics[scale=0.3]{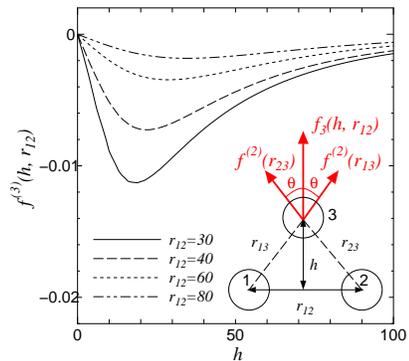}
\caption{
\label{3force}
The three-body force $f^{(3)}(h,r_{12})$ acting on the third disk 
as a function of $h$ defined in the inset.
Four curves are plotted for different values of $r_{12}$,
the inter-disk separation of disks 1 and 2.
The units of force and length are $k_BT/\lambda_B^{2}$ and $\lambda_B$, 
respectively.
}
\end{figure}

Thirdly, we consider a system with three disks in order to 
examine three-body interactions acting on charged colloidal disks.
Physical parameters were chosen identical to the two-disk
case.
The geometry of the three-disk system is shown in the inset of 
Fig.~\ref{3force}.
The third disk is located in the mid-plane of the first and the
second disks which have fixed separations $r_{12}$.
We calculated the total force $f_3(h,r_{12})$
acting on the third disk with varying $h$, the distance from
the axis connecting the first and second disks.
Since $f_3(h,r_{12})$ is always vertical because of the symmetry,
the vertical component of the three-body force $f^{(3)}(h,r_{12})$
acting on the third disk can be defined by
\begin{equation}
f^{(3)}(h,r_{12})\equiv f_3(h,r_{12})-f^{(2)}(r_{13})\cos\theta
- f^{(2)}(r_{23})\cos \theta,
\label{three}
\end{equation}
where $r_{ij}$ denotes the distance between the $i$th and $j$th disks,
and $\theta$ is defined as $\cos\theta = h/r_{13}$.
In Eq.~(\ref{three}), 
$r_{13}=r_{23}$ and $h=\sqrt{r_{13}^2-(r_{12}/2)^2}$ due to 
the symmetry of the geometry, and the pair force $f^{(2)}$ 
is given by the numerical results shown in Fig.~\ref{2force}.
Figure~\ref{3force} shows $f^{(3)}(h,r_{12})$ as a function of
$h$ with four different values of $r_{12}$.
Although we illustrate the numerical results only with $\xi=1.0$, 
we have carried out the same calculations also with $\xi=0.5$, $1.5$,
and $2.0$.
For the three-disk geometry examined here, 
all the curves tend to collapse onto each other within $1\%$ 
deviations.
It is seen in Fig.~\ref{3force} that the three-body force $f^{(3)}(h,r_{12})$ is
always attractive in this geometry. 
This tendency agrees well with the results in Ref.~\cite{C_Russ2002} 
where similar calculations have been performed by using FEM.
A similar tendency has been observed in 
microscopic MD~\cite{H_Lowen1998} and MC~\cite{JZ_Wu2000}
simulations as well as recent experiments~\cite{M_Brunner2004,J_Dobnikar2004}
for the same geometry in three-dimensional systems.

\begin{figure}
\includegraphics[scale=0.4]{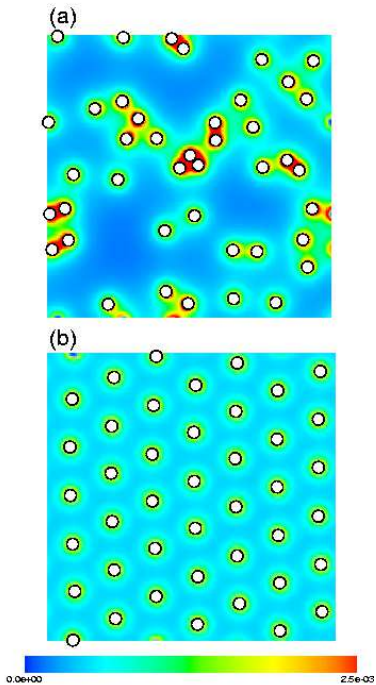}
\caption{
\label{42density}
Configurations of colloidal disks in the initial (a) and the 
final (b) states.
The temporal charge density
$(1-\phi({\bi r}))e\rho_c^{eq}({\bi r})$ is shown with a color map.
The scale of the color map is shown at the bottom 
in units of $e\lambda_B^{-3}$. 
}
\end{figure}

Finally, we performed a simple demonstration for the crystallization
of $N=42$ colloidal disks interacting each other.
Here, the system consists of $256\times 256$ grid points with the periodic
boundary condition and the
linear length of the system is $L=256$
while the other parameters are identical to those in the previous
cases of $N=2$ and $3$.
The positions of colloidal disks are followed by the steepest 
descent-type equation of motion
\begin{equation}
\zeta \frac{d {\bi R}_i}{dt}={\bi f}_i^{PP}+{\bi f}_i^{PS},
\label{motion_eq}
\end{equation}
which is obtained by substituting $d^2 {\bi R}_i/dt^2=0$, ${\bi
f}_i^R=0$, and ${\bi f}_i^{H}=-\zeta d{\bi R}_i/dt$ with the friction
constant $\zeta$ in Eq.~(\ref{motion}).
${\bi f}_i^{PP}\equiv -\partial E_{PP}/\partial {\bi R}_i$
is the force arising from the potential $E_{PP}$ acting directly 
between a pair
of colloidal disks.
We defined $E_{PP}$ as the repulsive part of the Lennard--Jones
potential,
$E_{PP}/k_BT=0.4\sum_{i=1}^{N-1}\sum_{j=i+1}^{N}
[(2a/|{\bi R}_i-{\bi R}_j|)^{12}-(2a/|{\bi R}_i-{\bi R}_j|)^{6}+1/4]$
truncated at the minimum distance $|{\bi R}_i-{\bi R}_j|=2^{7/6}a$.
In Fig.~\ref{42density}, we show snapshots of the
initial (a) and the final (b) configurations.
Starting from a non-overlapping random configuration shown in 
Fig.~\ref{42density}(a), 
the colloidal disks move simply to reduce the total free energy.
Eventually, the system attains a crystalline state with a
hexagonal close packed structure shown in Fig.~\ref{42density}(b) 
even at a very small packing fraction 
$\eta\equiv \pi a^2 N/L^2 \simeq 0.05$ of colloid without any
effective long-range interactions between colloidal disks. 
Similar crystalline structures have been observed in real experiments on 
charge-stabilized colloids~\cite{AP_Gast1998}.
It is worth mentioning the computational efficiency of our numerical method.
The demonstration shown in Fig~\ref{42density}, which required $600$
time steps, takes only one hour on a PC with a single Pentium4 2.8GHz CPU.

\section{concluding remarks}

A mesoscopic first-principle method are proposed for 
simulating charged colloidal dispersions.
In order to remove the numerical inefficiency due to the moving
boundary condition imposed on the colloid surface, 
a smoothed profile was introduced to represent the colloid-solvent 
interface.
In our method, the effects of counterions are 
considered within the framework of a density functional theory.
Furthermore, many-body effects among charged colloids are also
included properly.
We have examined the accuracy of the present method by changing
the interface thickness $\xi$ and found that the accuracy is satisfactory 
as far as the distance between colloidal disks is larger than 
$2(a+\xi)$.


Our final goal is to develop a simulation method applicable 
to dynamical problems of charged colloidal dispersions
such as electrophoresis where the coupling between hydrodynamics and
electrostatic interactions are crucial~\cite{Russel}.
In the present paper, we restricted our attentions only to static problems
with employing the adiabatic approximation, {\it i.e.},
$\rho_{\pm}({\bi r})$ follows instantaneously to the motions of 
the colloidal disks or particles.
This is OK for calculating stable colloidal structures in
the dispersions.
In the cases of dynamical problems, 
the time evolution of $\rho_{\pm}({\bi r})$ 
should be determined by coupling equations of hydrodynamics and 
thermal diffusion.
The PB equation is not appropriate for treating dynamical problems 
in which the counterion density
becomes anisotropic around a particle because of the friction between 
counterions and solvents.
Dipoles are induced if this happens, 
and thus interactions between colloids are no longer screened.
There appear long-range interactions between colloids, which must 
be important in many practical problems including electrophoresis 
for example.
Integration of the present method and the method for colloids in
Newtonian fluids \cite{Y_Nakayama2004} present promising approaches to
solve these cases, and efforts to this end are currently underway.

As this paper was being written for publication, we became
aware of a parallel effort by Kodama {\it et al.} \cite{H_Kodama2004}. 
While the omission of several details in their implementation make 
a detailed comparison between the two approaches impossible at this
time, it will be most useful in the future to
make a detailed comparison of the two methods both on 
efficiency and accuracy.

\begin{acknowledgments}

The authors are grateful to Dr. Y. Nakayama for helpful
discussions and comments.

\end{acknowledgments}


\end{document}